\documentclass[12pt,preprint]{aastex}

\shorttitle{X-ray Evolution of V1647 Ori}
\shortauthors{Kastner et al.}

\begin{document}


\title{V1647 Ori: The X-ray Evolution of a Pre-main Sequence Accretion Burst}  


\author{Joel H. Kastner\altaffilmark{1}, Michael
  Richmond\altaffilmark{1}, Nicolas Grosso\altaffilmark{2},
  David A. Weintraub\altaffilmark{3}, Theodore 
  Simon\altaffilmark{4}, Arne Henden\altaffilmark{5}, Kenji
  Hamaguchi\altaffilmark{6}, Adam Frank\altaffilmark{7}, Hideki
  Ozawa\altaffilmark{8}}    
\altaffiltext{1}{Rochester Institute of Technology, 54 Lomb Memorial Dr.,
  Rochester, NY 14623 (email: jhk@cis.rit.edu)}
\altaffiltext{2}{Laboratoire d'Astrophysique de Grenoble,
  Universit{$\acute{e}$} Joseph-Fourier, Grenoble, 38041, France}  
\altaffiltext{3}{Vanderbilt University, Nashville, Tennessee 37235, USA} 
\altaffiltext{4}{Institute for Astronomy, Honolulu, Hawaii 96822, USA} 
\altaffiltext{5}{US Naval Observatory, Flagstaff Station,
  P.O. Box 1149, Flagstaff, AZ 86002-1149}  
\altaffiltext{6}{NASA/Goddard Space Flight Center,
  Greenbelt, Maryland 20771; Universities Space Research Association,
  10211 Wincopin Circle, Suite 500, Columbia, Maryland 21044}  
\altaffiltext{7}{Department of Physics and Astronomy, University of
  Rochester, Rochester, NY 14627-0171} 
\altaffiltext{8}{Department of Earth and Space Science,
Graduate School of Science,
Osaka University, Osaka 560-0043, Japan}


\begin{abstract}

We present Chandra X-ray Observatory monitoring observations
of the recent accretion outburst displayed by the pre-main
sequence (pre-MS) star V1647 Ori. The X-ray observations
were obtained over a period beginning prior to outburst
onset in late 2003 and continuing through its apparent
cessation in late 2005, and demonstrate that the mean flux
of the spatially coincident X-ray source closely tracked the
near-infrared luminosity of V1647 Ori throughout its
eruption. We find negligible likelihood that the
correspondence between X-ray and infrared light curves over
this period was the result of multiple X-ray flares
unrelated to the accretion burst. The recent Chandra data
confirm that the X-ray spectrum of V1647 Ori hardened during
outburst, relative both to its pre-outburst state and to the
X-ray spectra of nearby pre-MS stars in the L1630 cloud. We
conclude that the observed changes in the X-ray emission
from V1647 Ori over the course of its 2003--2005 eruption
were generated by a sudden increase and subsequent decline
in its accretion rate. These results for V1647 Ori indicate
that the flux of hard X-ray emission from erupting low-mass,
pre-MS stars, and the duration and intensity of such
eruptions, reflect the degree to which star-disk magnetic
fields are reorganized before and during major accretion
events.

\end{abstract}




\section{Introduction}

The origin(s) of the luminous X-ray emission characteristic of
low-mass, pre-main sequence (pre-MS) stars remains the
subject of vigorous debate. For weak-lined (apparently
non-accreting) T Tauri stars, there is substantial evidence
that the X-rays are primarily the result of solar-like
coronal activity (Feigelson \& Montmerle 1999; Kastner et
al.\ 2004a). For certain actively accreting pre-MS stars,
however, the X-ray emission instead appears to arise as a
direct consequence of the accretion process itself (TW Hya,
Kastner et al.\ 2002, Stelzer \& Schmitt 2004; BP Tau,
Schmitt et al.\ 2005). These results cannot be easily 
reconciled with the fact that, for the Orion Nebula
Cluster, the X-ray luminosities of actively
accreting pre-MS stars are somewhat smaller than those of
non-accreting pre-MS stars (Flaccomio et al.\ 2003; Preibisch et
al.\ 2005).

Chandra X-ray Observatory and XMM-Newton observations of
V1647 Ori, obtained both before and just after this pre-MS
star in the L1630 dark cloud underwent a spectacular
optical/IR outburst (Briceno et al.\ 2004; Reipurth \& Aspin
2004; Walter et al.\ 2004; McGehee et al.\ 2004; Vacca,
Cushing, \& Simon 2004; Andrews et al.\ 2004; and references
therein), revealed a 
striking correspondence between the onset of the X-ray and
optical/IR eruptions of this pre-MS star (Kastner et al.\
2004b, hereafter K04; Grosso et al.\ 2005, hereafter
G05). These results serve as strong evidence that
high-energy emission from young stars can occur as a
consequence of high accretion rates. We have continued to
monitor V1647 Ori in X-rays with Chandra and XMM, and
present here the results of observations obtained beginning
about one year after the acquisition of the initial,
post-outburst X-ray data reported in K04 and G05.

\section{Observations and Data Reduction}

Observations of the L1630 field centered on V1647 Ori were
obtained on 2005 April 11, August 27, December 9 and 14, and
2006 May 1 with Chandra's Advanced CCD Imaging Spectrometer
in its imaging CCD readout configuration (ACIS-I). Exposure
times were 18.2, 19.9, 19.7, 18.1, and 21.7 ks,
respectively.  ACIS has a pixel size of 0.49$''$ and the
field of view of ACIS-I is $16'\times16'$; the
Chandra/ACIS-I combination is sensitive over the energy
range $\sim$0.3--10 keV. V1647 Ori was positioned at the
standard ACIS-I aimpoint on front-illuminated CCD I3 during
each exposure.  The data were subject to standard processing
by the Chandra X-Ray Center pipeline software (CIAO, V.\
3.1; CALDB, V.\ 3.0--3.2).  Count rates for V1647 Ori and
three nearby field X-ray sources (Table 1) were determined
by extracting source photons in the energy range 0.5-8.0 keV
(to limit background) within $2.5''$ radius circular
regions, and then subtracting the area-weighted background
count rate within a $20''$ radius off-source region. For the
2005 April and 2005 August datasets, in which V1647 Ori is
well detected, we also extracted X-ray light curves (Fig.\
1) and pulse height spectra (Fig.\ 2) within the $2.5''$
radius circular region centered on the position of the star.

\section{Results}

It is apparent from the count rates presented in Table 1
that the X-ray source associated with V1647 Ori generally
faded throughout the period of observations reported here,
to the extent that the source was not detected on 2005 Dec.\
and 2006 May 1; the detection on 2005 Dec.\ 14 is
marginal, with at most four source photons collected. The
three field X-ray sources --- all of which are associated
with pre-MS stars in the L1630 dark cloud --- were readily
detected (though variable in count rate) throughout this
same period (Table 1). It is furthermore apparent from Fig.\
1 that V1647 Ori remained variable on $\sim1$ hr timescales
during the $>1$ yr post-outburst X-ray observations reported
here; such rapid X-ray variability was also observed during
the first six months post-outburst (G05).

To assess whether the X-ray spectrum of V1647 Ori has
changed significantly since our previous post-outburst
observations (K04; G05), and to convert the V1647 Ori
Chandra/ACIS-I X-ray count rates to X-ray fluxes, we
performed spectral modeling with XSPEC (ver.\ 12). We
assumed the source spectrum consists of thermal plasma
emission (as represented by the MEKAL model; Liedahl et al.\
1995 and references therein) with a metallicity of 0.8
relative to solar and suffering intervening absorption
characterized by $N_H = 4.1\times10^{22}$ cm$^{-2}$
(G05). Under these assumptions, we find a source plasma with
$kT_X = 3.6 \pm 1.1$ keV best fits the 2005 April data
(Fig.\ 2). This value of $kT_X$ is well within the 90\%
confidence range of the ``hard'' X-ray spectral component
determined from modeling of Chandra data obtained in March
2004 (Fig.\ 2; K04) and XMM data obtained in 2004 April
(G05). The same plasma model is also consistent with the
rather noisy spectrum obtained in 2005 August (Fig.\ 2). We
therefore used this model ($N_H = 4.1\times10^{22}$
cm$^{-2}$, $kT_X = 3.6$ keV; ACIS-I3 energy conversion
factor ECF $=1.76\times10^{-11}$ erg cm$^{-2}$ count$^{-1}$) to
calculate observed X-ray fluxes (and X-ray flux upper
limits) from the ACIS count rates of V1647 Ori for all five
observations reported here, as well as for the post-outburst
2004 March observations. The measured median photon energies
(Table 1) confirm that the V1647 Ori X-ray source remained
somewhat hard for at least one year post-outburst, relative
both to its pre-outburst X-ray emission and to other nearby,
pre-MS sources in L1630. The median energies provide
marginal evidence 
that the X-ray spectrum of V1647 Ori softened after 2005
August, so we adopt the estimated pre-outburst value $kT_X =
0.86$ keV (K04) in converting the 2005 December and 2006 May
ACIS-I3 count rate and upper limit to X-ray
flux and flux upper limit,  respectively (adopting
the ECF calculated from the 2005 April data would increase
both flux values by a factor $\sim2$).

These X-ray flux results, and contemporaneous near-infrared
fluxes, are compiled in Fig.\ 3, which also includes previously
reported near-infrared and X-ray fluxes. For consistency
with K04 and G05 --- and with the infrared photometric data,
which are not corrected for extinction ---
values of X-ray flux included in Fig.\ 3 are not corrected for
intervening absorption. For the four Chandra and XMM
exposures of duration $\ge20$ ks in which the mean X-ray flux is $F_X
> 10^{-14}$ erg cm$^{-2}$ s$^{-1}$, we indicate the range of X-ray
flux as measured in 5 ks time bins. This Figure reveals that,
subsequent to April 2005 (J.D. 2,453,400), V1647 Ori steeply declined
in near-infrared luminosity, such that the object had returned to near
pre-outburst levels by late 2005.  The XMM and Chandra/ACIS-I
observations reveal a steady decline in mean X-ray flux from the
object over this time period. We next consider whether the
two declines are physically related.

\section{Discussion}

\subsection{Striking correspondence or mere coincidence?}

Flux variations of a factor $\sim10$ were measured in the
longer-duration XMM observations of V1647 Ori in outburst
(Fig.\ 3). While the detailed X-ray variations of V1647 Ori
observed in these longer exposures (e.g., G05) do not appear
typical of X-ray flares from more evolved, low-mass, pre-MS
stars, it is important to assess the likelihood that the
apparent striking correspondence between the X-ray and
near-infrared light curves of V1647 Ori over the course of
its outburst resulted from random X-ray flaring during the
optical/IR eruption.

Consider a simplified ``short-term'' ($<100$ ks duration)
flaring model in which the X-ray source could be in either
of two states --- flaring (``high'', mean $F_X >
10^{-14}$ erg cm$^{-2}$ s$^{-1}$) or quiescent (``low'', mean 
$F_X < 10^{-14}$ erg cm$^{-2}$ s$^{-1}$) --- with equal
probability, during any
given Chandra or XMM observation. For 10 random samplings of
such a bimodal source, the probability that we would
observe the specific sequence of mean source states displayed in
Fig.\ 3 --- low prior to J.D. 2,453,000, high between
J.D. 2,453,000 and 2,453,700, low again after J.D. 2,453,700
--- is 1 in $2^{10}$, or $\sim0.1$\%. We also constructed a
somewhat more sophisticated Monte Carlo flare model, in
which the source is characterized by a specific combination
of flare duration and flare duty cycle, each of which are
adjustable parameters. The source state was then determined
at irregular temporal sampling intervals matching those of
the Chandra and XMM observations. We find that, over a wide
range of both flare duration (10 ks to 100 ks) and duty
cycle (10\% to 90\%), the number of instances in which the
model high vs.\ low state sequence matches that of the
observations is no more than 0.11\% of the total number of
model sequences.  Hence, it is highly unlikely that the apparent
coincidence of strongly enhanced and then sharply declining
X-ray flux is the chance superposition of shorter-duration
X-ray flaring on the longer-term near-infrared flux increase
and decline.

\subsection{Classification of V1647 Ori}

Herbig (1989) described eruptive behavior among low-mass,
pre-MS stars as either FU Orionis-like (``FUor''; see review
by Kenyon \& Hartmann 1996) or EX Lupi-like (``EXor''). Such
sustained optical/IR outbursts from pre-MS stars are
generally interpreted as episodes of rapid accretion (e.g.,
Herbig et al.\ 2001; Hartmann \& Kenyon 1996). There has
been much discussion in the recent literature (e.g., Aspin
et al.\ 2006; Vig et al.\ 2006; Ojha et al.\ 2006; Muzerolle
et al.\ 2005) as to which accretion burst ``class'' V1647 Ori might
belong. Based on the near-infrared monitoring results
presented in Fig.\ 3 (see also Aspin et al.\ 2006), it is
now apparent that the outburst that began in late 2003 and
had ended by late 2005 was an event of much shorter duration
than those of ``classical'' FUors, which typically last
several decades. The only other known outburst of V1647 Ori
during the past 100 yr, which occurred in 1966-1967, was of
similar or perhaps somewhat shorter duration (Aspin et al.).
It therefore appears V1647 Ori cannot be considered an FUor
and, hence, joins the ranks of the candidate EXors --- which
already comprise a rather diverse set of objects in terms of
outburst amplitudes, durations, and frequencies. Such a
broad spectrum of variability behavior ultimately hints at
the presence of a wide range in the rates of accretion onto
individual pre-MS stars, as well as a wide range in the rate
of {\it change} of these accretion rates.

\subsection{X-rays and accretion}

The increase and subsequent decrease in the infrared flux
from V1647 Ori (Fig.\ 3) is evidently entirely the result of
variations in accretion luminosity, as there has been
minimal change in extinction toward the object during its
outburst (Reipurth \& Aspin 2004; Gibb et al.\
2006). Specifically, the accretion rate is
inferred to have increased from a pre-outburst level of a
few $\times10^{-7}$ $M_\odot$ yr$^{-1}$ to $\sim10^{-5}$
$M_\odot$ yr$^{-1}$ at outburst onset (Vacca et al.\ 2004;
Muzerolle et al.\ 2005). It follows that the rapid decline
in the optical/IR luminosity of of V1647 Ori in late 2005
was caused by a similarly precipitous drop in
accretion rate to near pre-outburst levels. The measured
Br-gamma line flux from V1647 Ori dropped by a factor
$\sim10$ between 2005 March and 2006 January (Gibb et al.\
2006; Simon, unpublished), supporting the interpretation
that the accretion rate has declined sharply over this
period.

Hence, just as the sudden onset of bright X-rays from V1647
Ori was closely coupled to its initial accretion burst
(K04), the recent drop in the X-ray luminosity of V1647 Ori
is most likely directly coupled to a rapid decline in
stellar accretion rate. The contrast between the relatively
small amplitude short-term near-infrared variability of
V1647 Ori during its 2003--2005 outburst and its rapid
(hour-timescale), large-amplitude X-ray variability over
this same period further suggests that whereas the
near-infrared continuum is diagnostic of large-scale changes
in disk mass and temperature structure (that, in turn,
reflect changes in the time-averaged accretion rate;
Muzerolle et al.\ 2005), the X-rays trace small-scale changes in
the star-disk magnetic field configuration that result in
the release of magnetic energy.

These results for V1647 Ori, like the detection of enhanced
X-ray emission from XZ Tau during an optical/IR outburst
from that star (Giardino et al.\ 2006), strengthen the link
between pre-MS accretion and X-ray emission. On the other
hand, the recent optical outburst of the EXor candidate
V1118 Ori was accompanied by a {\it softening} of its X-ray
spectrum and only modest changes in X-ray luminosity (Audard
et al.\ 2005); and simultaneous optical and X-ray
observations of the Orion Nebula Cluster obtained over a
$\sim2$-week period indicate that, on timescales of hours to
days, optical and X-ray variability is largely uncorrelated
(Stassun et al.\ 2006). Neither the V1118 Ori nor ONC
optical/IR outbursts were as intense as that of V1647 Ori,
and the ONC optical flares observed by Stassun et al.\ are,
of course, of far shorter duration. 

The foregoing suggests that the
duration and intensity of a pre-MS stellar eruption reflects
the degree to which the magnetic fields connecting star to
disk are reorganized before and during a major accretion event.
The degree of reorganization (i.e., the intensity of
associated magnetic reconnection events) then likely determines
whether such an accretion event is accompanied by the
release of hard X-rays, as opposed to soft X-rays indicative
of accretion shocks (e.g., Kastner et al.\ 2002). The
unusually hard X-ray spectrum of FU Ori itself, with its
prominent Fe line emission (Skinner, Briggs, \&  G{\"u}del 2006),
may indicate that this system remains in a volatile state of
ongoing star-disk magnetic field reconnection.  Additional,
long-term, simultaneous X-ray and near-infrared monitoring
of eruptive pre-MS stars is essential to establish whether
sustained (EXor- or FUor-like) accretion bursts are commonly
accompanied by enhanced hard X-ray emission.

\acknowledgements{The authors thank the referee for helpful
  suggestions. This research was supported by the National
  Aeronautics and Space Administration via Chandra Awards
  GO5--6003X and GO6--7004X issued to R.I.T. by the Chandra
  X-ray Obsevatory Center, which is operated on behalf of
  NASA by Smithsonian Astrophysical Observatory under
  contract NAS8--03060.}


\clearpage

\begin{figure}
\includegraphics[scale=1.0,angle=0]{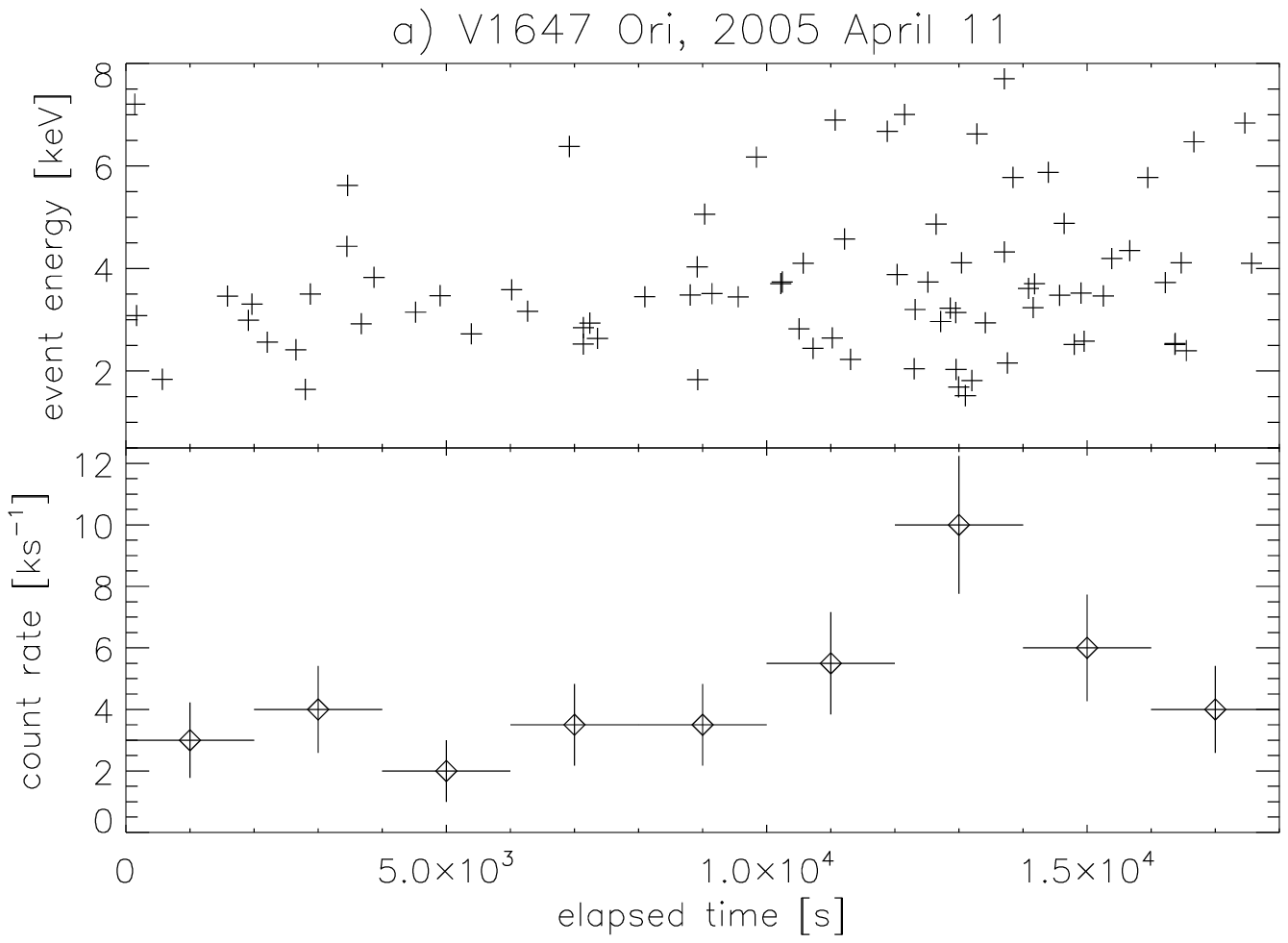}
\includegraphics[scale=1.0,angle=0]{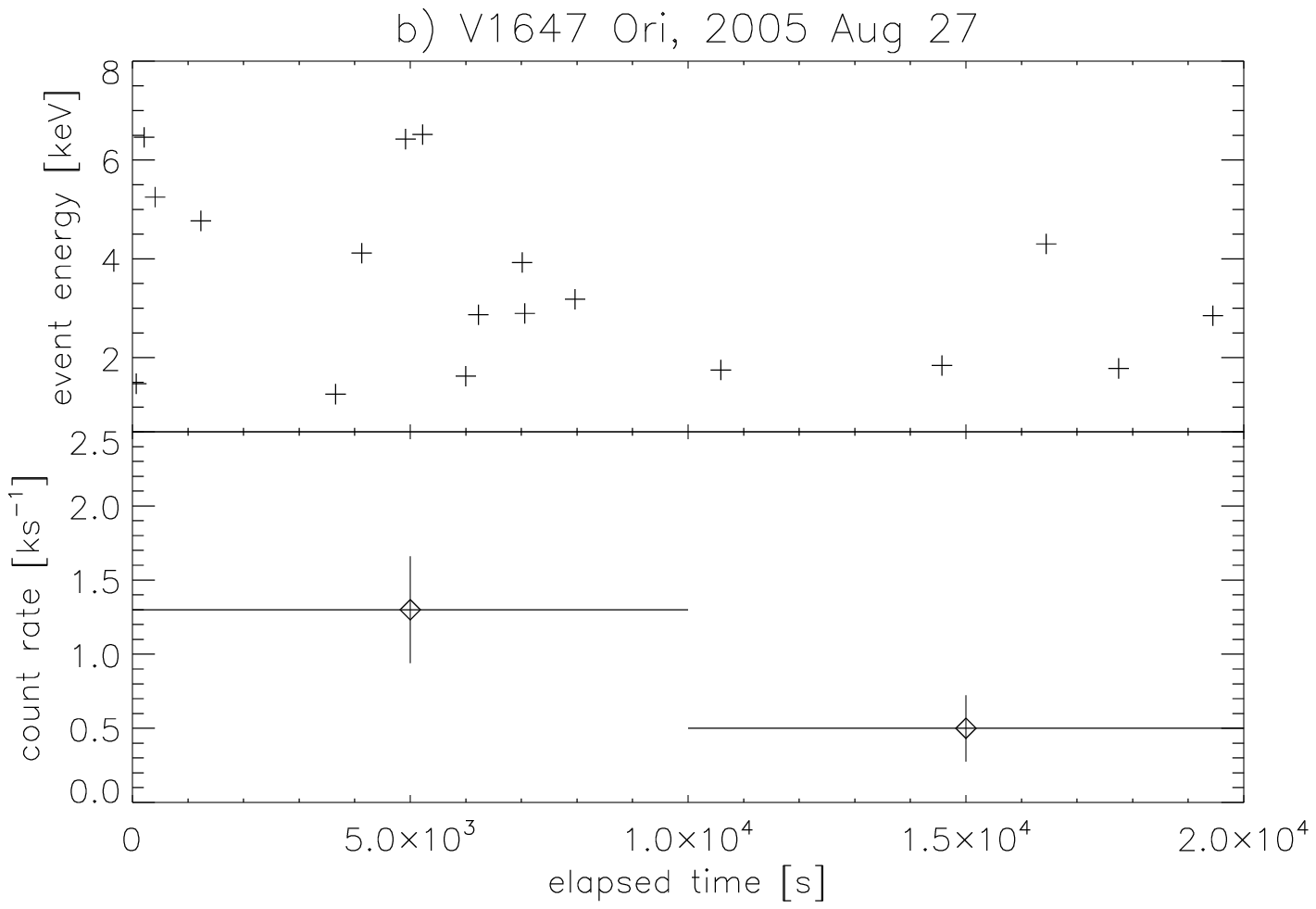}
\caption
{Chandra/ACIS-I3 photon scatter plots (energy vs.\ time) and
  light curves for V1647 Ori observations of (a) 2005 Apr.\ 11
  and (b) 2005 Aug.\ 27. In the light curves (lower panels),
  horizontal error bars indicate the widths of the 
  time bins, and vertical error bars indicate 1 $\sigma$
  uncertainties in count rate. During
  each observation, variations in count rate 
  on timescales of $\sim1-3$ hr are apparent.} 
\end{figure}

\clearpage

\begin{figure}
\begin{center}
\includegraphics[scale=1.5,angle=0]{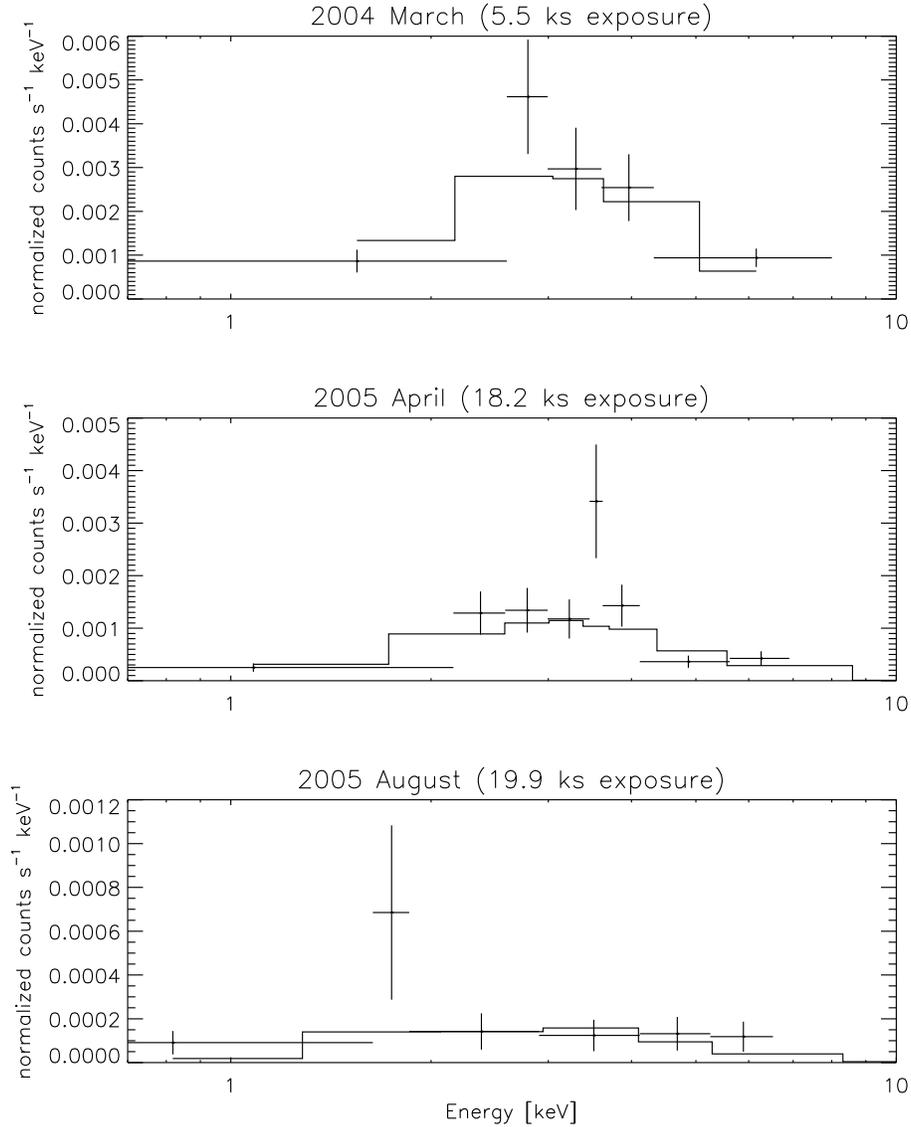}
\end{center}
\caption
{Chandra/ACIS spectra (crosses) obtained for V1647 Ori in
  2004 March 7
  (top), 2005 April (middle), and 2005 August (bottom). The
  same absorbed thermal plasma model ($N_H =
  4.1\times10^{22}$ cm$^{-2}$, $kT_X = 3.6$ keV) is
  renormalized to obtain the best fit and
  overlaid on each spectrum (histogram). See \S 3.}
\end{figure}

\clearpage

\begin{figure}
\begin{center}
\includegraphics[scale=0.75,angle=0]{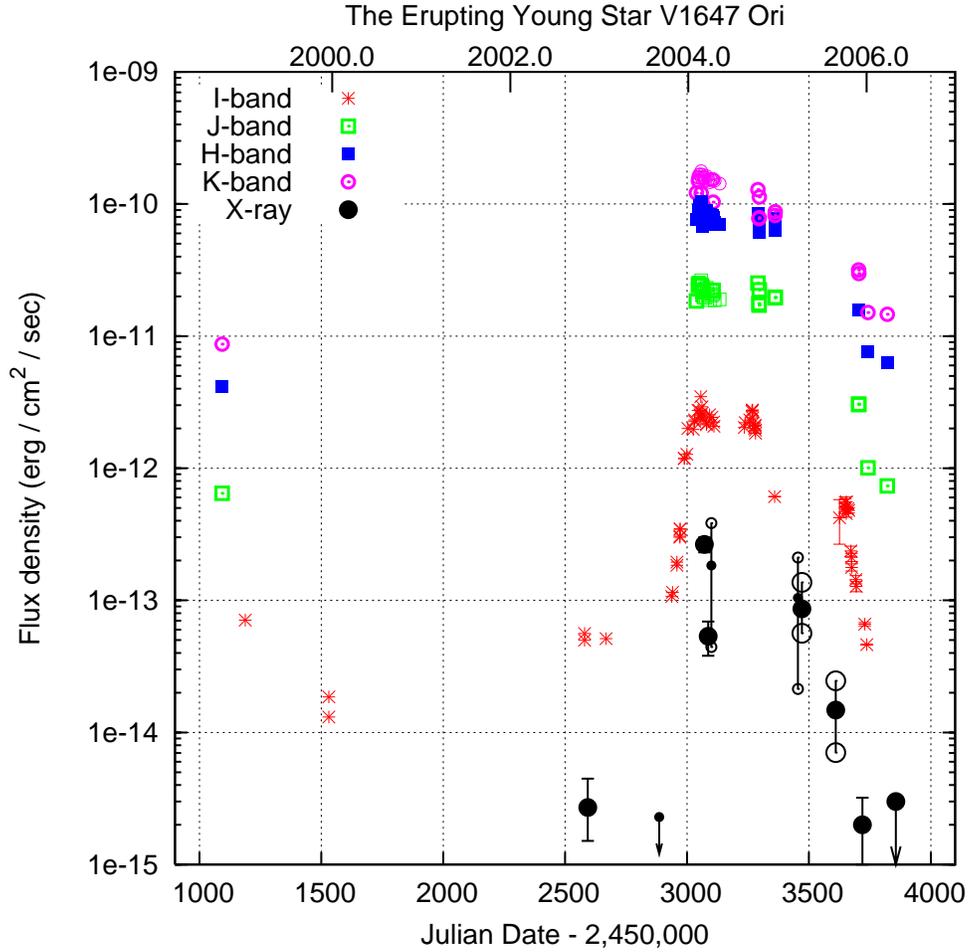}
\end{center}
\caption
{ Near-IR and X-ray light curves of V1647 Ori covering the
  period from 1998 to May 2006. Calendar year is indicated along
  the top of the Figure. Data prior to J.D. 2,453,200
  were reported previously in K04 and G05. Subsequent
  near-infrared measurements 
  were obtained by Kospal et al.\ (2006), Henden
  (unpublished), Fedele \& van den Ancker (2006), Venkat \&
  Anandarao (2006), and Vig et al.\ (2006), and with the RIT
  Observatory 16-inch Meade reflector and SBIG CCD camera
  (single I-band measurement with larger error bar). The
  X-ray flux measurements
  were obtained by CXO (larger circles) 
  and XMM (smaller circles; XMM data obtained
  J.D. 2,453,454 are from a 90 ks exposure [Grosso et
  al. 2006, in prep.]). The vertical lines connecting
  the pairs of open circles denote the observed ranges of
  X-ray flux, as measured in 5 ks time bins, during
  individual exposures; the filled
  circles along these lines indicate observed mean X-ray fluxes during these
  same exposures. Error bars on other X-ray measurements are
  1 $\sigma$. The CXO upper limit at J.D. 2,453,714 has been
  omitted, for clarity. Neither the infrared nor X-ray flux
  measurements have been corrected for extinction or absorption.
}
\end{figure}

\clearpage

\begin{table}
\setlength{\tabcolsep}{0.025in}
\begin{center}
\footnotesize
\caption{ACIS$^a$ count rates and median photon energies
  of L1630 X-ray sources}  
\begin{tabular}{lccccccccc}
\hline
\hline
       &  & \multicolumn{2}{c}{V1647 Ori} &
\multicolumn{2}{c}{LkH$\alpha$ 301} & \multicolumn{2}{c}{CXO J054618.8-000537} &
      \multicolumn{2}{c}{CXO J054611.6-000627} \\ 
 Obs.\ Date  & J.D.
& rate & $E^b$ & rate & $E^b$ & rate & $E^b$ & rate & $E^b$ \\ 
& & (ks$^{-1}$) & (keV) & (ks$^{-1}$) & (keV) & (ks$^{-1}$) &
 (keV) & (ks$^{-1}$) & (keV) \\
\hline
2002 Nov.\ 14 & 2452593 & 0.21$\pm0.06$ & 1.8$\pm0.5$ &  8.3$\pm$0.4  & 1.4$\pm0.1$ &
  32.2$\pm$0.8   & 1.4$\pm0.1$    &  2.0$\pm$0.2  & 1.6$\pm0.1$ \\ 
2004 Mar.\ 7  & 2453072 & 10.7$\pm$1.4  & 3.6$\pm0.3$ &  8.9$\pm$1.4  & 1.6$\pm0.2$ &
  26.2$\pm$2.3   & 1.4$\pm0.1$    &  5.8$\pm$ 1.1 & 1.7$\pm0.2$ \\
2004 Mar.\ 22 & 2453087 & 2.2$\pm$0.7   & 2.0$\pm$0.4 &  5.7$\pm$1.1  & 1.8$\pm0.3$ &
  25.1$\pm$2.3   & 1.5$\pm0.1$    &  1.4$\pm$0.5  & 1.1$\pm0.5$ \\
2005 Apr.\ 11 & 2453472 & 4.9$\pm$0.5   & 3.5$\pm$0.1 &  2.1$\pm$0.4  & 2.0$\pm0.2$ & 
  50.0$\pm$2.0   & 1.9$\pm0.1$    &  1.4$\pm$0.4  & 2.1$\pm0.2$ \\
2005 Aug.\ 27 & 2453610 & 0.9$\pm$0.2   & 3.0$\pm$0.7 &  1.3$\pm$0.3  & 1.5$\pm0.1$ &
  15.6$\pm$1.0   & 1.6$\pm0.1$    &  1.6$\pm$0.3  & 1.8$\pm0.3$ \\
2005 Dec.\ 9  & 2453714 & $<0.25^c$     & \nodata     &  3.7$\pm$0.4  & 1.6$\pm0.1$ &
  12.7$\pm$1.1   & 1.4$\pm0.1$    &  0.56$\pm$0.08 & 1.5$\pm1.1$ \\
2005 Dec.\ 14 & 2453719 & 0.20$\pm$0.12 & 1.7$\pm0.3$ &  4.6$\pm$0.5  & 1.6$\pm0.1$ &
  17.1$\pm$1.0   & 1.6$\pm0.1$    &  1.1$\pm$0.3  & 1.7$\pm0.2$ \\ 
2006 May 1    & 2453857 & $<0.3^c$      & \nodata     &  15.3$\pm$0.8 & 1.8$\pm0.1$ &
  18.5$\pm$0.9   & 1.6$\pm0.1$    &  0.4$\pm$0.15 & 1.9$\pm1.0$ \\
\hline
\end{tabular}
\footnotesize
\end{center}
\noindent
Notes: \\
a) ACIS-S in 2002 and 2004 observations, ACIS-I in 2005--2006; these
four sources were located on 
a front-illuminated CCD in the 2002 and 2005--2006 observations, 
and on a back-illuminated CCD in the 2004 observations. \\
b) Median energy of source photons, with uncertainties
calculated via the half sample method (Babu \& Feigelson 
1996). \\
c) 3$\sigma$ upper limit.
\end{table}


\begin{thebibliography}{}


\bibitem[]{425} Andrews, S.M., Rothberg, B., \& Simon,
  T. 2004, ApJ, 610, L45 

\bibitem[]{427} Aspin, C., Barbieri, C., Boschi, F., diMille, F.,
  Rampazzi, F., Reipurth, B., \& Tsvetkov, M. 2006, AJ, in press

\bibitem[]{430} Audard, M.,   G{\"u}del, M., Skinner, S. L., Briggs,
  K. R., Walter, F. M., Stringfellow, G., Hamilton, R. T., \&
  Guinan, E. F. 2005, ApJ, 635, L81

\bibitem[]{434} Babu, G. J., \& Feigelson,
  E. D. 1996, ``Astrostatistics'' (Chapman \& Hall), pp.\ 96--97

\bibitem[]{437} Briceno, C. et al. 2004, ApJ, 606, L123

\bibitem[]{439} Fedele, D., \& van den Ancker, M. E. 2006, CBET, 428, 1

\bibitem[]{441} Feigelson, E.D., \& Montmerle, T. 1999, ARAA, 37, 363


\bibitem[]{445} Flaccomio, E., Damiani, F., Micela, G., Sciortino, S.,
  Harnden, F. R., Jr., Murray, S. S., \& Wolk, S. J. 2003, ApJ, 582, 398


\bibitem[]{450} Giardino, G., Favata, F., Silva, B., Micela, G.,
  Reale, F., \& Sciortino, S. 2006, A\&A, in press (astro-ph/0603630)

\bibitem[]{453} Gibb, E. L., Rettig, T. W., Brittain, S. D., Wasikowski, D.,
Simon, T., Vacca, W. D., Cushing, M. C., \& Kulesa, C. 2006,
ApJ, 641, 383

\bibitem[]{457} Grosso, N., Kastner, J., Ozawa H., Richmond M., Simon T.,
  Weintraub D.A., Hamaguchi K., \& Frank A. 2005, A\&A, 438, 
  159 (G05)

\bibitem[]{461} Hartmann, L., \& Kenyon, S.J. 1996, ARAA, 34, 207



\bibitem[]{467} Herbig, G. 1989, in Proceedings of ``ESO Workshop on Low Mass
  Star Formation and Pre-main Sequence Objects,'' B. Reipurth, ed.
 (Munich: ESO), p.\ 233

\bibitem[]{471} Herbig, G., Aspin, C., Gilmore, A. C., Imhoff, C. L., \&
  Jones, A. F. 2001, PASP, 113, 1547 

\bibitem[]{474} Kastner, J.H., Huenemoerder, D.P., Schulz,
N.S., Canizares, C.R., \& Weintraub, D.A. 
2002, ApJ, 567, 434.

\bibitem[]{478} Kastner, J.H., 
Huenemoerder, D.P., Schulz,
N.S., Canizares, C.R., Li, J., \& Weintraub, D.A. 
2004a, ApJ, 605, L49

\bibitem[]{483} Kastner, J.H., et al.
  2004b, Nature,
  430, 429 (K04) 


\bibitem[]{492} Kospal, A., Abraham, P., Acosta-Pulido, J., Csizmadia, Sz.,
 Eredics, M., Kun, M., Racz, M. 2006, IBVS, in press (astro-ph/0511733)

\bibitem[]{495} Liedahl, D.A., Osterheld, A.L., \& Goldstein, W.H. 1995, ApJL, 438,
115


\bibitem[]{500} McGehee, P., Smith, J.A., Henden, A.A.,
Richmond, M.W., Knapp, G.R., Finkbeiner, D.P., Ivezic, Z., \&
Brinkmann, J. 2004, ApJ, 616, 1058 


\bibitem[]{504} Muzerolle, J., Megeath S.T., Flaherty K.M., Gordon K.D.,
  Rieke G.H., Young E.T., \& Lada C.J. 2005, ApJ, 620, L107 


\bibitem[]{509} Preibisch, Th., et al. 2005, ApJS, 160, 557

\bibitem[]{511} Ojha, D.K., et al. 2006, MNRAS, 368, 825

\bibitem[]{513} Reipurth, B., \& Aspin, C. 2004, ApJ, 606, L119

\bibitem[]{515} Schmitt, J.H.M.M., Robrade, J., Ness, J.-U., Favata, F., \&
  Stelzer, B. 2005, A\&A, 432, L35 

\bibitem[]{518} Skinner, S. L., Briggs, K. R., \&  G{\"u}del, M. 2006,
  ApJ, 643, 995



\bibitem[]{525} Stassun, K., van den Berg, M., Feigelson, E., Flaccomio,
  E. 2006, ApJ, in press (astro-ph/0606079)

\bibitem[]{528} Stelzer, B., \& Schmitt, J.H.M.M. 2004, A\&A, 418, 687

\bibitem[]{} Vacca, W.D., Cushing, M.C., \& Simon, T. 2004,
  ApJ, 609, L29

\bibitem[]{530} Venkat, V., \& Anandarao, B. G. 2006, IAU Circular 8694

\bibitem[]{532} Vig, S., Ghosh, S.K., Kulkarni, V.K., Ojha, D.K. 2006,
  A\&A, 446, 1021

\bibitem[]{} Walter, F.M., Stringfellow, G.S., Sherry, W.H., \&
Field-Pollatou, A. 2004, AJ, 128, 1872


\end{thebibliography}
\end{document}